\documentclass[a4paper,fleqn,usenatbib]{mnras}

\usepackage{newtxtext,newtxmath}

\usepackage[T1]{fontenc}
\usepackage{ae,aecompl}

\usepackage{graphicx}	
\usepackage{amsmath}	
\usepackage{amssymb}	

\title[Solar chemical composition of the ICM]{Solar chemical composition in the hot gas of cool-core ellipticals, groups, and clusters of galaxies}

\author[F. Mernier et al.]{
F. Mernier,$^{1,2,3}$\thanks{E-mail: mernier@caesar.elte.hu}
N. Werner$^{1,4,5}$
J. de Plaa,$^{3}$
J. S. Kaastra,$^{3,6}$
A. J. J. Raassen,$^{3}$
L. Gu,$^{7}$
\newauthor
J. Mao,$^{3,6}$ 
I. Urdampilleta$^{3,6}$
and A. Simionescu$^{8}$
\\
$^{1}$MTA-E\"otv\"os University Lend\"ulet Hot Universe Research Group, P\'azm\'any P\'eter s\'et\'any 1/A, Budapest, 1117, Hungary\\
$^{2}$Institute of Physics, E\"otv\"os University, P\'azm\'any P\'eter s\'et\'any 1/A, Budapest, 1117, Hungary\\
$^{3}$SRON Netherlands Institute for Space Research, Sorbonnelaan 2, 3584 CA Utrecht, The Netherlands\\
$^{4}$Department of Theoretical Physics and Astrophysics, Faculty of Science, Masaryk University, Kotl\'a\v{r}sk\'a 2, Brno, 611 37, Czech Republic \\
$^{5}$School of Science, Hiroshima University, 1-3-1 Kagamiyama, Higashi-Hiroshima 739-8526, Japan \\
$^{6}$Leiden Observatory, Leiden University, P.O. Box 9513, 2300 RA Leiden, The Netherlands\\
$^{7}$RIKEN Nishina Center, 2-1 Hirosawa, Wako, Saitama 351-0198, Japan \\
$^{8}$Institute of Space and Astronautical Science (ISAS), JAXA, 3-1-1 Yoshinodai, Chuo-ku, Sagamihara, Kanagawa 252-5210, Japan
}

\date{Accepted 2018 July 18. Received 2018 July 18; in original form 2018 June 27}

\pubyear{2018}

\begin{document}
\label{firstpage}
\pagerange{\pageref{firstpage}--\pageref{lastpage}}
\maketitle

\begin{abstract}
The hot intracluster medium (ICM) pervading galaxy clusters and groups is rich in metals, which were synthesised by billions of supernovae and have accumulated in cluster gravitational wells for several Gyrs. Since the products of both Type Ia and core-collapse supernovae -- expected to explode over different time scales -- are found in the ICM, constraining accurately the chemical composition of these hot atmospheres can provide invaluable information on the history of the enrichment of large-scale structures. Recently, \textit{Hitomi} observations reported solar abundance ratios in the core of the Perseus cluster, in tension with previous \textit{XMM-Newton} measurements obtained for 44 cool-core clusters, groups, and massive ellipticals (the CHEERS sample). In this work, we revisit the CHEERS results by using an updated version of the spectral code used to fit the data (\textsc{spexact} v3), the same as was used to obtain the \textit{Hitomi} measurements. Despite limitations in the spectral resolution, the average Cr/Fe and Ni/Fe ratios are now found to be remarkably consistent with unity and in excellent agreement with the \textit{Hitomi} results. Our updated measurements suggest that the solar composition of the ICM of Perseus is a common feature in nearby cool-core systems.
\end{abstract}

\begin{keywords}
supernovae: general -- ISM: abundances -- galaxies: clusters: intracluster medium -- X-rays: galaxies: clusters
\end{keywords}




\section{Introduction}
\label{sec:intro}

Since the hot, X-ray emitting haloes -- or intracluster medium (ICM) -- pervading galaxy clusters, groups, and massive ellipticals are in collisional ionisation equilibrium (CIE), abundances of the chemical elements they retain over Gyrs can be robustly measured. Among them, oxygen (O), neon (Ne), and magnesium (Mg) are mainly produced by core-collapse supernovae (SNcc). Chromium (Cr), manganese (Mn), iron (Fe), and nickel (Ni), on the other hand, are mainly produced by Type Ia supernovae (SNIa). Intermediate elements such as silicon (Si), sulfur (S), argon (Ar), calcium (Ca) originate from both SNIa and SNcc \citep[for recent reviews, see e.g.][]{2008SSRv..134..337W,2013ARA&A..51..457N}. 

Interestingly, these two types of supernovae originate from very different progenitors with different lifetimes and, consequently, should not enrich their surroundings in the same way and with the same delays. Whereas SNcc result from the explosion of a massive star and occur a few million years after the formation of their progenitor, SNIa are thought to occur with a substantial delay, i.e. when a white dwarf (WD) gains enough material from a companion object (either a main-sequence star or another degenerate core remnant) to ignite explosive nucleosynthesis. Consequently, accurate measurements of the chemical composition of the ICM (or of any other astrophysical system) provide invaluable clues to understand its enrichment, as well as its subsequent evolution \citep[for previous studies, see e.g.][]{1996ApJ...466..686M,2002A&A...381...21F,2005ApJ...620..680B,2007A&A...465..345D,2007ApJ...667L..41S}.

Recently, we compiled deep \textit{XMM-Newton} EPIC and RGS observations of 44 nearby cool-core ellipticals, galaxy groups, and clusters (the CHEERS\footnote{CHEmical Enrichment Rgs Sample} catalogue) in order to measure accurately the O/Fe, Ne/Fe, Mg/Fe, Si/Fe, S/Fe, Ar/Fe, Ca/Fe, Cr/Fe, Mn/Fe, and Ni/Fe ratios and to derive a complete abundance pattern, representative of the nearby ICM \citep[][hereafter Paper I]{2016A&A...592A.157M}. This pattern was found to be consistent with solar values for the seven former ratios while, on the contrary, Cr/Fe, Mn/Fe, and Ni/Fe were measured higher than solar. In a second paper \citep[][hereafter Paper II]{2016A&A...595A.126M}, we compared the measured X/Fe abundance ratios with recent or commonly used SNIa and SNcc yield models, in order to provide reliable constraints on SN explosions and/or their progenitors. In particular, we showed that an accurate determination of Ni/Fe is extremely important as it may dramatically alter our astrophysical interpretations of the ICM enrichment.

Some of these results, however, appeared to be in tension with recent measurements of the Perseus cluster core obtained with the \textit{Hitomi} observatory \citep[][hereafter H17]{2017Natur.551..478H}. Unlike our super-solar values measured for Cr/Fe, Mn/Fe, and Ni/Fe in Paper I, the excellent spectral resolution offered by the micro-calorimeter SXS showed, instead, a chemical composition that is remarkably close to that of the Solar neighbourhood for all the investigated ratios. These discrepancies have been recently confirmed by \citet[][hereafter S18]{2018arXiv180600932S}, who revisited the Perseus abundance ratios by combining \textit{Hitomi} SXS and \textit{XMM-Newton} RGS instruments. Although these three studies (Paper I; H17; S18) are currently the state-of-the-art of our knowledge of the chemical composition of the ICM, the discrepancies noted above may find various origins, since the observations were performed on different systems (44 systems vs. Perseus only), with different instruments (CCDs vs. micro-calorimeter) and with different spectral codes (see below). Two questions particularly arise: (i) To which extent are CCD instruments reliable to constrain abundance ratios? (ii) Is the chemical composition of Perseus rather unique, or do most other clusters exhibit a solar chemical composition as well?

Spectral codes, like \textsc{apec} \citep{2012ApJ...756..128F} and \textsc{spex} \citep[][]{1996uxsa.conf..411K} have had major updates in preparation for \textit{Hitomi}. Although the codes still differ \citep{2018PASJ...70...12H}, they have converged considerably in recent years.
Between 2016 and 2017, however, a major update of the atomic data included in the spectral fitting package \textsc{spex} has been publicly released. Whereas H17 used the new version of the code, Papers I and II appeared beforehand. It was shown recently that this update has a major impact on measuring the Fe abundance in groups and ellipticals \citep{2018MNRAS.478L.116M} and could thus potentially affect the abundance of other elements \citep[see also figure~9 in][]{2017A&A...603A..80M}. By essence, the estimated abundances of a CIE plasma depend on the input atomic calculations in the spectral model that is used. Therefore, comparing the CHEERS results with the best measurements of the Perseus cluster (H17; S18) in a consistent way is essential for setting the most accurate constraints on the chemical composition of the ICM.

In this Letter, we explore the effects of such model improvements on our previously measured abundance ratios in the CHEERS sample (Paper I). We assume $H_0$ = 70 km s$^{-1}$ Mpc$^{-1}$, $\Omega_m$ = 0.3, and $\Omega_\Lambda$= 0.7. Unless otherwise stated, the error bars are given within a 68\% confidence interval. All the abundances mentioned in this work are given with respect to the proto-solar values (referred to as "solar" for convenience) derived from \citet{2009LanB...4B...44L}.

\section{Methods}
\label{sec:methods}

\subsection{The sample}
\label{sec:the_sample}

The sample, the data reduction, and the spectral analysis and strategies are all described in detail in Paper I. Like our present work, that previous study focused on the \textit{XMM-Newton} observations of 44 nearby ($z < 0.1$) cool-core clusters, groups, and ellipticals, all being part of the CHEERS project \citep[see also][]{2015A&A...575A..38P,2017A&A...607A..98D}. The main criterion of the sample is the detection of the \ion{O}{viii} emission line measured by the RGS instrument with >5$\sigma$ of significance. In this way, we ensure selecting clusters with prominent metal lines in their cores. This allows a robust determination of most of the abundances also with the EPIC instruments.

As explained in Paper I, by selection the 24 hottest systems of the sample ($kT > 1.7$ keV; "clusters") are investigated within $0.2 r_{500}$ while the remaining 20 cooler systems  ($kT < 1.7$ keV; "groups/ellipticals") could only be studied within $0.05 r_{500}$. The only exception is the very nearby elliptical galaxy M\,87, whose central temperature is about $\sim$2 keV but whose $0.2 r_{500}$ limit stands beyond the EPIC field of view. This system is thus studied within $0.05 r_{500}$. In Paper I, we showed that adopting these different extraction radii has negligible impact on our final results.

\subsection{Reanalysis of our data}\label{sec:SPEX2vs3}

Since 1996, the original \textsc{mekal} code, used to model thermal plasmas \citep{1972A&A....20..215M,1985A&AS...62..197M,1986A&AS...65..511M}, has been developed independently within the \textsc{spex} spectral fitting package \citep{1996uxsa.conf..411K} and gradually improved. Up to the version 2.06, the code made use of an atomic database and a collection of routines that are all referred to \textsc{spexact} (\textsc{spex} Atomic Code and Tables) v2.
Since 2016, however, a major update (version 3.04, hereafter \textsc{spexact} v3) has been released on both the atomic database and the corresponding routines. 
As described in \citet{2017A&A...607A..98D} and \citet{2018MNRAS.478L.116M}, the main changes include the incorporation of $\sim$400 times more transitions (including higher principal quantum numbers for H-like and He-like iones) as well as (i) a more realistic treatment of the radiative recombination emission \citep{2016A&A...587A..84M}, (ii) improvements in collisional ionisation calculations \citep{2017A&A...601A..85U}, and (iii) updates of collisional (de-)excitation rates, radiative transition probabilities, auto-ionization, and dielectronic recombination rates.

In Paper I, the O/Fe and Ne/Fe ratios were already corrected to their \textsc{spexact} v3 estimates; therefore, there is no need to reconsider them further and we devote this Letter to the EPIC measurements of Mg/Fe, Si/Fe, S/Fe, Ar/Fe, Ca/Fe, Cr/Fe, Mn/Fe, and Ni/Fe. The general fitting procedure, including (i) the estimation of the hydrogen column density $n_\text{H}$, (ii) the treatment of the EPIC background, and (iii) the approach of refitting K-shell lines locally, is described extensively in Paper I. The only exception regarding (iii) is M\,87, which show significantly different "full band" and "local" Fe values, and for which we adopt the latter (since its Fe-L complex may suffer from unexpected biases, such as a complex temperature structure). In addition, we account  for the multi-temperature state of the plasma by adopting the approach described in \citet{2018MNRAS.478L.116M}. In short, three thermal components are assumed, each with free temperatures. The emission measures of the hotter and cooler component, however, are tied to half that of the main component (of temperature $kT_\text{mean}$) in order to approximate a Gaussian temperature distribution (the \texttt{gdem} model; see e.g. Paper I) with reasonable computing resources.


\section{Results}
\label{sec:results}

\begin{figure}
	\includegraphics[trim=0.9cm 0.6cm 1.5cm 1.9cm, clip=true,width=\columnwidth]{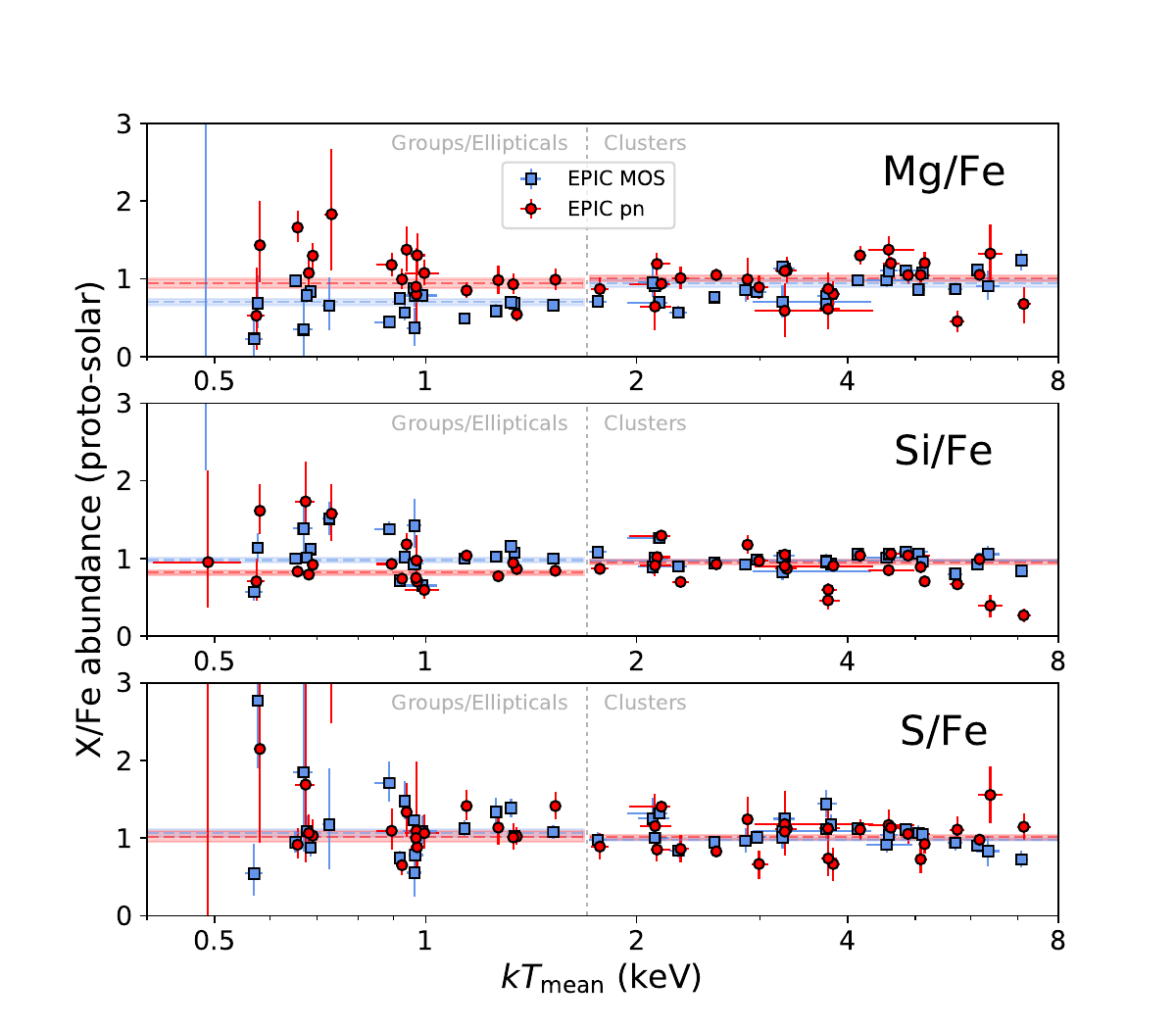}
    \caption{Abundance ratios (\textit{top:} Mg/Fe; \textit{middle:} Si/Fe; \textit{bottom:} S/Fe) as a function of the mean temperature of the CHEERS systems (using \textsc{spexact} v3). EPIC MOS and pn measurements are shown separately. The vertical dotted line ($kT_\text{mean} = 1.7$ keV) separates the "groups/ellipticals" and the "clusters" subsamples. The blue and red horizontal dashed lines (and filled areas) indicate the mean values (and errors) averaged over these two sub-samples for MOS and pn, respectively (see text).}
    \label{fig:kT_XFe}
\end{figure}

In Fig.~\ref{fig:kT_XFe}, we report the Mg/Fe, Si/Fe, and S/Fe ratios of the CHEERS systems as a function of their $kT_\text{mean}$. We choose to report the EPIC MOS and pn results individually because these instruments are known to have slight but significant discrepancies in their best-fit parameters \citep[][]{2015A&A...575A..30S,2015A&A...575A..37M}. It clearly appears that these ratios remain very similar in the full range of $kT_\text{mean}$ considered here. In fact, when averaging these ratios over the "clusters" and "groups/ellipticals" subsamples (Fig.~\ref{fig:kT_XFe}, dashed lines and filled areas), we find that except a moderate ($\sim$26\%) decrease of Mg/Fe from "clusters" to "groups/ellipticals" in the MOS measurements, the other measurements show either <1$\sigma$ consistent mean ratios, or very limited (i.e. less than 15\%) differences, with no systematic trend. Since Mg, Si, and S may be considered as reliable tracers of SNcc products while Fe originates predominantly from SNIa (e.g. Paper II), the similar chemical composition in all our systems strongly suggests that SNIa and SNcc enrich the hot atmosphere of clusters, groups, and ellipticals with the same relative importance. Similar conclusions were obtained by \citet{2009A&A...508..565D}, \citet{2007A&A...465..345D}, and in Paper I (albeit with more restricted samples and/or outdated spectral codes).

\begin{figure}
	\includegraphics[trim=0.3cm 0.7cm 0.5cm 0.4cm, clip=true,width=\columnwidth]{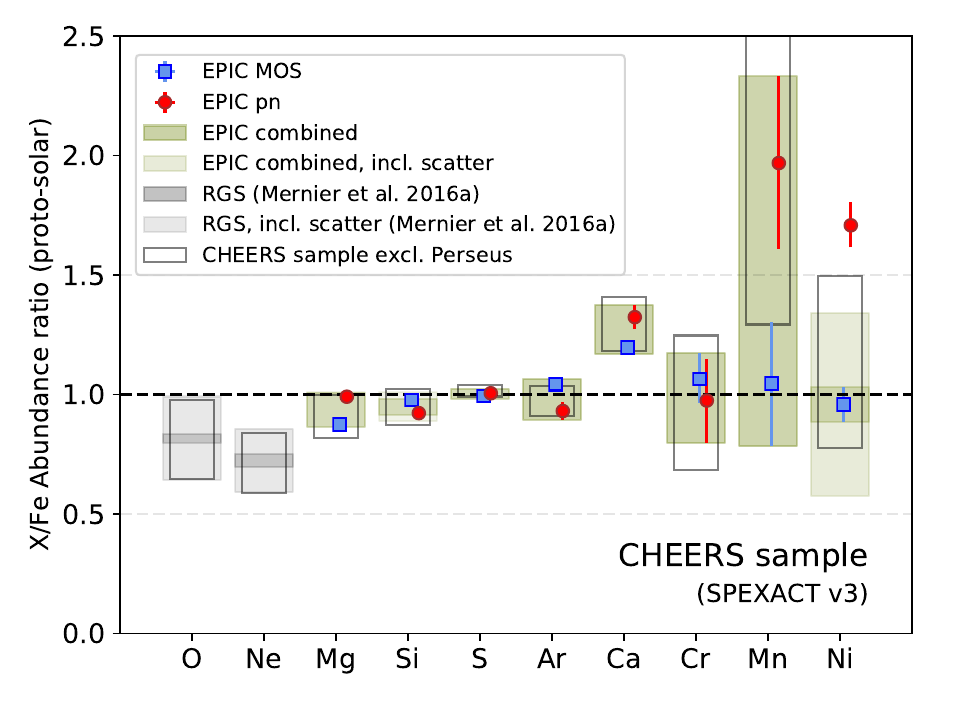}
    \caption{Average abundance ratios re-estimated from the entire CHEERS sample (using \textsc{spexact} v3), for EPIC MOS and pn separately (blue and red data points), then adopting conservative limits accounting for the MOS-pn discrepancies (green darker boxes) and possible additional scatter (green lighter boxes). For Ni/Fe, only the MOS measurements are considered in these conservative limits (see text). The O/Fe and Ne/Fe ratios (grey boxes) are adopted from Paper I (already corrected from \textsc{spexact} v3). The same conservative limits obtained when excluding Perseus from the sample are shown by the black contours.}
    \label{fig:abun_CHEERS}
\end{figure}

The remarkable similarity of the X/Fe ratios in all these systems also allows us to average them over the entire sample in order to obtain a combined abundance pattern, representative of the (nearby cool-core) ICM as a whole. These CHEERS average abundance ratios are shown in Fig.~\ref{fig:abun_CHEERS} for MOS and pn independently (along with the RGS measurements of O/Fe and Ne/Fe from Paper I). Because in most cases the statistical uncertainties are lower than the respective MOS-pn discrepancies, it is very important to account for the latter in our total uncertainties. In order to be conservative, for each ratio we consider the two extreme values reached by the individual MOS and pn measurements (and their 1$\sigma$ statistical uncertainties) as the total MOS-pn cross-calibration uncertainties (green darker boxes in Fig.~\ref{fig:abun_CHEERS}). This approach allows to define mean values for the X/Fe ratios and their associated conservative limits ($\sigma_\text{cons}$; including the statistical uncertainties and, when applicable, the MOS-pn discrepancies), as provided in Table~\ref{table:systematics_SPEX3}.

\begin{table}
\begin{centering}
\caption{Average abundance ratios re-estimated from the CHEERS sample (using \textsc{spexact} v3), as well as their systematic and total uncertainties. An absence of value ($-$) means that no further uncertainty was required.}             
\label{table:systematics_SPEX3}
 \setlength{\tabcolsep}{9pt}
\begin{tabular}{c| c| c c c}        
\hline \hline                
Element & Mean value& $\sigma_\text{cons}$ & $\sigma_\text{int}$ & $\sigma_\text{tot}$ \\    

\hline                        
 O/Fe & $0.817$ & $0.018$ & $0.174$ & $0.175$ \\
 Ne/Fe & $0.724$ & $0.028$ & $0.130$ & $0.133$ \\
Mg/Fe & $0.937$ & $0.071$ & $0.013$ & $0.072$ \\
Si/Fe & $0.949$ & $0.034$ & $0.051$ & $0.061$ \\
S/Fe & $1.004$ & $0.021$ & $-$ & $0.021$ \\
Ar/Fe & $0.980$ & $0.085$ & $-$ & $0.085$ \\
Ca/Fe & $1.272$ & $0.103$ & $-$ & $0.103$ \\
Cr/Fe & $0.986$ & $0.188$ & $-$ & $0.188$ \\
Mn/Fe & $1.557$ & $0.774$ & $-$ & $0.774$ \\
Ni/Fe & $0.959$ & $0.073$ & $0.375$ & $0.382$ \\

\hline                                   
\end{tabular}
\par\end{centering}
\end{table}

The Ni/Fe ratio deserves some extra attention. In particular, the large discrepancy between MOS and pn measurements suggests that this ratio is very sensitive to the instrumental background. In fact, in pn spectra we note the presence of two strong and unstable fluorescent lines, at $\sim$7.5 keV (Ni K$\alpha$) and $\sim$8.0 keV (Cu K$\alpha$). Since these instrumental lines are partly blended with the Ni-K complex from the source, they strongly bias the pn measurements of the Ni abundance. On the contrary, no instrumental feature is reported in MOS spectra within this band. This explains the inconsistent values between MOS and pn, which previously led to large systematic error bars (figure~6 right of Paper I). For this reason, in the rest of this Letter we only rely on the MOS values of Ni/Fe.

In addition, individual measurements may also have an (limited) intrinsic scatter, due to the slight differences in the chemical histories of each system. This additional uncertainty, $\sigma_\text{int}$, is evaluated as described in Paper I. In short, the combined MOS+pn measurements are fit with a constant. If $\chi^2/\text{d.o.f.} > 1$, we increment $\sigma_\text{int}$ that we add in quadrature to all the individual measurements until $\chi^2/\text{d.o.f.}$ reaches unity. This scatter (green lighter boxes in Fig.~\ref{fig:abun_CHEERS}) is then added in quadrature to the MOS-pn discrepancies (or, for the Ni/Fe ratio, to the MOS results only) to obtain our final uncertainties, $\sigma_\text{tot}$ (see Table~\ref{table:systematics_SPEX3}).

Our CHEERS abundance pattern is found to be remarkably consistent with the solar ratios. However, because (i) Perseus is the brightest object of the CHEERS sample and (ii) the abundance pattern of that system is already known to be solar (H17; S18), one might wonder how this particular system weights the results of our entire sample. In Fig.~\ref{fig:abun_CHEERS} (black contours), we show how the conservative limits presented above change when the Perseus measurements are excluded from the sample. The excellent consistency between the two patterns clearly illustrates that, like Perseus, most other systems also show a chemical composition that is surprisingly close to solar.

\section{Comparison with previous measurements}
\label{sec:Ni_bias}

\begin{figure}
	\includegraphics[trim=0.3cm 0.7cm 0.5cm 0.4cm, clip=true,width=\columnwidth]{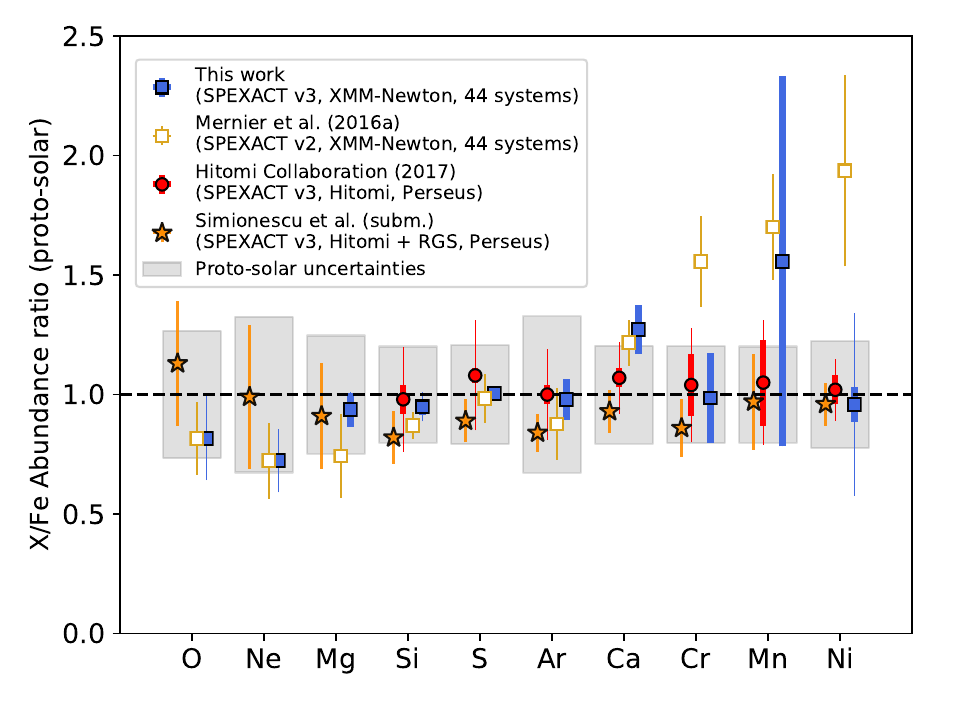}
    \caption{Comparison of our updated (\textsc{spexact} v3) averaged abundance ratios (blue squares; thick error bars include statistical uncertainties and MOS-pn discrepancies while thin error bars include additional scatter) with recent results from the literature. For comparison, current uncertainties on the solar ratios \citep{2009LanB...4B...44L} are shown by the grey boxes.}
    \label{fig:abun_CHEERS_comparison}
\end{figure}

We have shown that, when applying the latest atomic database to the CHEERS sample, the chemical composition of the ICM is found to be remarkably similar to that of the Solar neighbourhood. As shown in Fig.~\ref{fig:abun_CHEERS_comparison}, this result is in excellent agreement with the recent measurements of several abundance ratios in the Perseus core by \textit{Hitomi} (H17; see also S18), although the measurements were done on different systems (the CHEERS sample vs. Perseus) and with different instruments (EPIC CCDs vs. SXS micro-calorimeter). The solar Cr/Fe and Ni/Fe ratios found in this work differ from those in Paper I. We conclude that the previous discrepancies in Cr/Fe and Ni/Fe reported by H17 (see their figure~2) were due to the use of different spectral code versions rather than the moderate spectral resolution of EPIC. These solar abundance ratios were also obtained in Perseus using the up-to-date version of \textsc{apec}, with limited differences compared to \textsc{spexact} v3 \citep[S18; see also][]{2018PASJ...70...12H}.

\begin{figure}
	\includegraphics[trim=0.2cm 0cm 0.2cm 0.6cm, clip=true,width=\columnwidth]{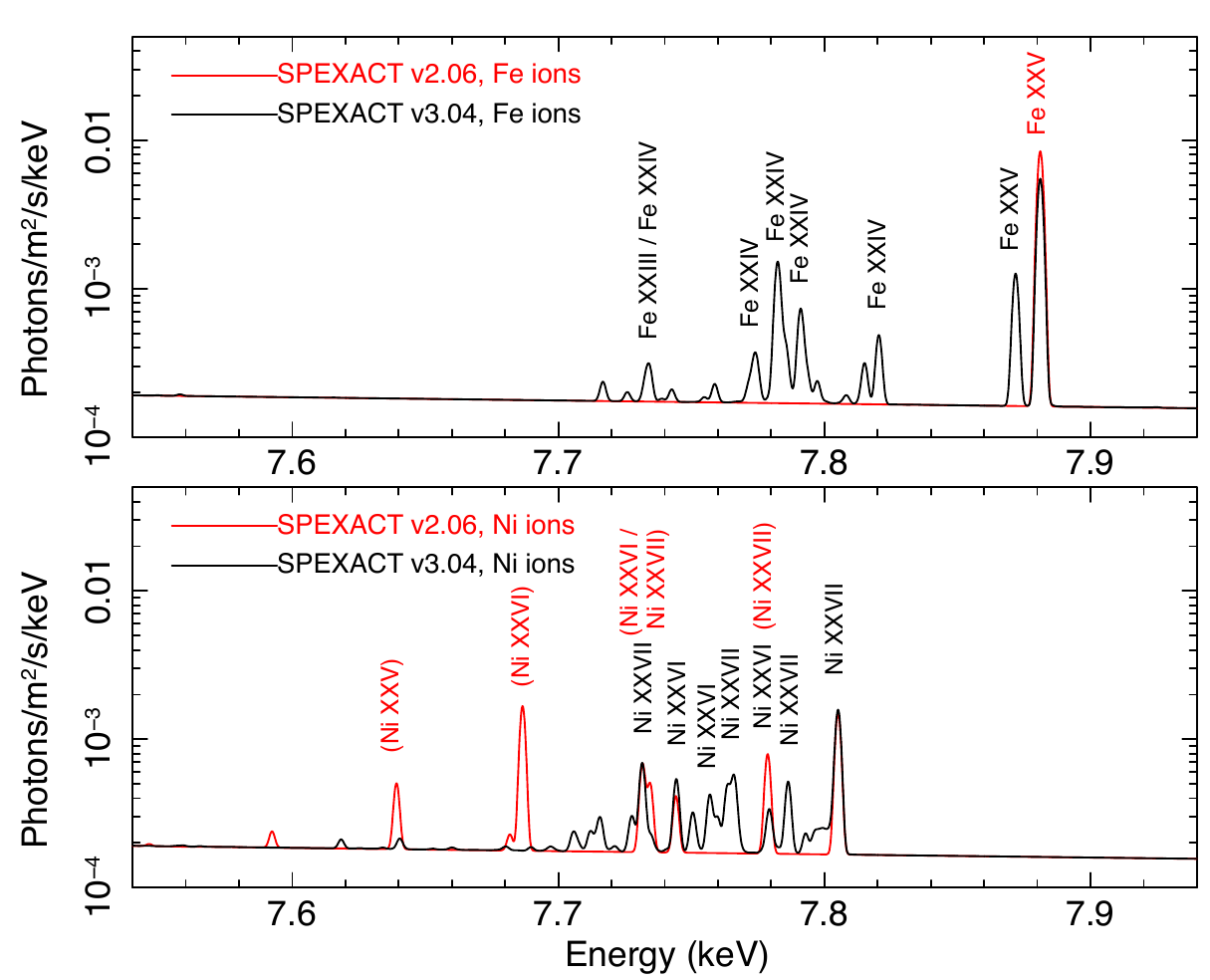}
    \caption{Comparison between \textsc{spexact} v2 and \textsc{spexact} v3 for a $kT = 3$ keV plasma, zoomed on the Ni-K complex ($\sim$7.5--8 keV). The transitions of the Fe and Ni ions are shown separately in the upper and lower panels, respectively.}
    \label{fig:Ni_models}
\end{figure}

Among the \textsc{spexact} v2--v3 differences seen in Fig.~\ref{fig:abun_CHEERS_comparison}, the Ni/Fe ratio is the most striking. To better understand the reasons for such a decrease, we show in Fig.~\ref{fig:Ni_models} the CIE emission calculated successively with these two versions for a moderately hot plasma ($kT = 3$ keV, solar abundances). 
Within the Ni-K energy band ($\sim$7.5--8 keV), only Fe and Ni ions produce emission lines, and we separate the transitions of these two elements in the upper and lower panels of Fig.~\ref{fig:Ni_models}, respectively.

Although the emissivities of many Ni lines have been notably revised with the latest update of \textsc{spexact}, we note that the overall equivalent width (EW) of all the Ni-K lines remain comparable between v2 and v3 (Fig.~\ref{fig:Ni_models}, lower panel). On the contrary, while \textsc{spexact} v2 includes only one Fe transition in the Ni-K complex (\ion{Fe}{xxv} at $\sim$7.88 keV), \textsc{spexact} v3 shows that many more Fe lines (mostly from \ion{Fe}{xxiii}, \ion{Fe}{xxiv}, and \ion{Fe}{xxv}) contaminate this energy band (Fig.~\ref{fig:Ni_models}, upper panel). Assuming that \textsc{spexact} v3 reproduces realistically all the transitions that contribute to the Ni-K bump observed with EPIC, the incorrectly high Ni/Fe ratio measured in Paper I can be naturally explained. Indeed, in order to compensate for the total EW of the unaccounted Fe lines in the Ni-K complex, \textsc{spexact} v2 incorrectly raised the Ni parameter until it fully fits the Ni-K bump.

Unlike the other ratios measured with EPIC, Cr/Fe and Mn/Fe mentioned in Paper I were not calculated with \textsc{spexact} v2 but with a precursor of \textsc{spexact} v3.00. Between that intermediate version and the one used in this work (v3.04), significant transitions have been included and/or updated. In particular, \ion{Cr}{xxii} transitions significantly contribute to the total EW of the Cr-K complex, but were only incorporated later on, in the up-to-date version. The absence of such transitions in previous models explain why the Cr/Fe was significantly overestimated. Similarly, the EW of Mn-K transitions have been revised higher, which explains why the MOS average Mn/Fe ratio is now remarkably consistent with the solar value. Its pn counterpart, however, is measured significantly larger than solar (though with large error bars). The origin of such a discrepancy is, unfortunately, difficult to identify. Unlike the case of Ni, the Mn-K band does not contain instrumental lines that may affect one specific detector. Therefore, Mn cannot be well constrained with EPIC instruments (see also discussion in S18).

Another ratio of interest is Ca/Fe. Compared to Paper I, this ratio remains essentially unchanged and appears somewhat in tension with the (lower) value reported by S18. This discrepancy was already pointed out in Perseus only (S18; see their figure~9); nevertheless this ratio is left unchanged when discarding Perseus from our analysis (Fig.~\ref{fig:abun_CHEERS}). Interestingly, nucleosynthesis models are currently unable to reproduce the \textit{XMM-Newton} measurements of Ca/Fe \citep[][Paper II]{2007A&A...465..345D}. Several possibilities have been investigated, among which alternative SNIa models reproducing the spectral features of the Tycho supernova \citep{2007A&A...465..345D} or a significant contribution of Ca-rich gap transients to the ICM enrichment \citep[][Paper II]{2014ApJ...780L..34M}. However, the significantly lower Ca/Fe measurements of Perseus in both \textit{Hitomi} and \textit{Suzaku} suggest an issue that could be specific to the EPIC instruments.

\section{Implications for the chemical enrichment of the central ICM}
\label{sec:implications}

More than demonstrating the relative reliability of CCD instruments in deriving ICM abundances, our results strongly suggest that a solar chemical composition is a feature common to most (if not all) nearby cool-core systems, and not specific to Perseus only.

As already discussed by S18, such a "simple" chemical composition of the central ICM is not trivial to explain. Stellar populations of massive early-type galaxies, which are often found in the centre of groups and clusters, exhibit super-solar $\alpha$/Fe ratios, probably associated to relatively short star formation timescales \citep{2014ApJ...780...33C}. In fact, if the bulk of SNIa explode and enrich their surroundings with a significant delay compared to the end of the parent star formation, their products are not efficiently incorporated in the stellar population we see today (S18). While enrichment by SNIa dominate in the brightest cluster galaxy (BCG) -- where SNcc are rarely, if ever, observed, these $\alpha$-enriched stars might also enrich the surrounding ICM via their stellar winds \citep[e.g.][]{2009A&A...493..409S}. It would be a surprising coincidence, however, that for nearly all the CHEERS systems these two sources of ICM enrichment would compensate each other to reach exactly the solar ratios within their current uncertainties.
In addition, the remarkable uniformity in the chemical composition of the ICM from the very core to the outskirts \citep[][]{2015ApJ...811L..25S,2017A&A...603A..80M} may seriously question the role of the BCG in the central enrichment of clusters and groups. 

In summary, our results strengthen the "ICM solar composition paradox" already reported by H17 and S18, and extend it to a large number of cool-core clusters, groups, and ellipticals. Interesting ways of exploring this paradox in the future are to constrain (i) the redshift evolution of $\alpha$/Fe ratios and (ii) the abundance of other (in particular odd-$Z$) elements in the ICM. While sodium (Na) or aluminium (Al) are crucial to constrain the initial metallicities of SNcc progenitors \citep{2013ARA&A..51..457N}, lighter elements such as carbon (C) or nitrogen (N) are mostly enriched via asymptotic giant branch stars, and their C/Fe and N/Fe ratios are not necessarily expected to be solar \citep[][Mao et al. submitted]{2006A&A...459..353W,2011A&A...531A..15G}. Extending abundance studies to all these unexplored ratios will definitely help to understand how and when the ICM has been enriched.
This is out of the scope of this Letter, as deeper observations with micro-calorimeter instruments onboard future missions (e.g. \textit{XARM}, \textit{Athena}) are required.

\section*{Acknowledgements}

We thank the referee for useful feedback that helped to improve this Letter. This work was supported by the Lend\"ulet LP2016-11 grant awarded by the Hungarian Academy of Sciences. This work is partly based on the \textit{XMM-Newton} AO-12 proposal ``\emph{The XMM-Newton view of chemical enrichment in bright galaxy clusters and groups}'' (PI: de Plaa), and is a part of the CHEERS (CHEmical Evolution Rgs cluster Sample) collaboration. This work is based on observations obtained with \textit{XMM-Newton}, an ESA science mission with instruments and contributions directly funded by ESA member states and the USA (NASA). The SRON Netherlands Institute for Space Research is supported financially by NWO, the Netherlands Organisation for Scientific Research.




\bibliographystyle{mnras}
\bibliography{Letter_II_ratios}

\begin{thebibliography}{}
\makeatletter
\relax
\def\mn@urlcharsother{\let\do\@makeother \do\$\do\&\do\#\do\^\do\_\do\%\do\~}
\def\mn@doi{\begingroup\mn@urlcharsother \@ifnextchar [ {\mn@doi@}
  {\mn@doi@[]}}
\def\mn@doi@[#1]#2{\def\@tempa{#1}\ifx\@tempa\@empty \href
  {http://dx.doi.org/#2} {doi:#2}\else \href {http://dx.doi.org/#2} {#1}\fi
  \endgroup}
\def\mn@eprint#1#2{\mn@eprint@#1:#2::\@nil}
\def\mn@eprint@arXiv#1{\href {http://arxiv.org/abs/#1} {{\tt arXiv:#1}}}
\def\mn@eprint@dblp#1{\href {http://dblp.uni-trier.de/rec/bibtex/#1.xml}
  {dblp:#1}}
\def\mn@eprint@#1:#2:#3:#4\@nil{\def\@tempa {#1}\def\@tempb {#2}\def\@tempc
  {#3}\ifx \@tempc \@empty \let \@tempc \@tempb \let \@tempb \@tempa \fi \ifx
  \@tempb \@empty \def\@tempb {arXiv}\fi \@ifundefined
  {mn@eprint@\@tempb}{\@tempb:\@tempc}{\expandafter \expandafter \csname
  mn@eprint@\@tempb\endcsname \expandafter{\@tempc}}}

\bibitem[\protect\citeauthoryear{{Baumgartner}, {Loewenstein}, {Horner}  \&
  {Mushotzky}}{{Baumgartner} et~al.}{2005}]{2005ApJ...620..680B}
{Baumgartner} W.~H.,  {Loewenstein} M.,  {Horner} D.~J.,   {Mushotzky} R.~F.,
  2005, \mn@doi [\apj] {10.1086/427158}, \href
  {http://adsabs.harvard.edu/abs/2005ApJ...620..680B} {620, 680}

\bibitem[\protect\citeauthoryear{{Conroy}, {Graves}  \& {van Dokkum}}{{Conroy}
  et~al.}{2014}]{2014ApJ...780...33C}
{Conroy} C.,  {Graves} G.~J.,   {van Dokkum} P.~G.,  2014, \mn@doi [\apj]
  {10.1088/0004-637X/780/1/33}, \href
  {http://adsabs.harvard.edu/abs/2014ApJ...780...33C} {780, 33}

\bibitem[\protect\citeauthoryear{{De Grandi} \& {Molendi}}{{De Grandi} \&
  {Molendi}}{2009}]{2009A&A...508..565D}
{De Grandi} S.,  {Molendi} S.,  2009, \mn@doi [\aap]
  {10.1051/0004-6361/200912745}, \href
  {http://adsabs.harvard.edu/abs/2009A%26A...508..565D} {508, 565}

\bibitem[\protect\citeauthoryear{{Finoguenov}, {Matsushita}, {B{\"o}hringer},
  {Ikebe}  \& {Arnaud}}{{Finoguenov} et~al.}{2002}]{2002A&A...381...21F}
{Finoguenov} A.,  {Matsushita} K.,  {B{\"o}hringer} H.,  {Ikebe} Y.,   {Arnaud}
  M.,  2002, \mn@doi [\aap] {10.1051/0004-6361:20011477}, \href
  {http://adsabs.harvard.edu/abs/2002A%26A...381...21F} {381, 21}

\bibitem[\protect\citeauthoryear{{Foster}, {Ji}, {Smith}  \&
  {Brickhouse}}{{Foster} et~al.}{2012}]{2012ApJ...756..128F}
{Foster} A.~R.,  {Ji} L.,  {Smith} R.~K.,   {Brickhouse} N.~S.,  2012, \mn@doi
  [\apj] {10.1088/0004-637X/756/2/128}, \href
  {http://adsabs.harvard.edu/abs/2012ApJ...756..128F} {756, 128}

\bibitem[\protect\citeauthoryear{{Grange}, {de Plaa}, {Kaastra}, {Werner},
  {Verbunt}, {Paerels}  \& {de Vries}}{{Grange}
  et~al.}{2011}]{2011A&A...531A..15G}
{Grange} Y.~G.,  {de Plaa} J.,  {Kaastra} J.~S.,  {Werner} N.,  {Verbunt} F.,
  {Paerels} F.,   {de Vries} C.~P.,  2011, \mn@doi [\aap]
  {10.1051/0004-6361/201016187}, \href
  {http://adsabs.harvard.edu/abs/2011A%26A...531A..15G} {531, A15}

\bibitem[\protect\citeauthoryear{{Hitomi Collaboration} et~al.,}{{Hitomi
  Collaboration} et~al.}{2017}]{2017Natur.551..478H}
{Hitomi Collaboration} et~al., 2017, \mn@doi [\nat] {10.1038/nature24301},
  \href {http://adsabs.harvard.edu/abs/2017Natur.551..478H} {551, 478}

\bibitem[\protect\citeauthoryear{{Hitomi Collaboration} et~al.,}{{Hitomi
  Collaboration} et~al.}{2018}]{2018PASJ...70...12H}
{Hitomi Collaboration} et~al., 2018, \mn@doi [\pasj] {10.1093/pasj/psx156},
  \href {http://adsabs.harvard.edu/abs/2018PASJ...70...12H} {70, 12}

\bibitem[\protect\citeauthoryear{{Kaastra}, {Mewe}  \&
  {Nieuwenhuijzen}}{{Kaastra} et~al.}{1996}]{1996uxsa.conf..411K}
{Kaastra} J.~S.,  {Mewe} R.,   {Nieuwenhuijzen} H.,  1996, in {Yamashita} K.,
  {Watanabe} T.,  eds, UV and X-ray Spectroscopy of Astrophysical and
  Laboratory Plasmas. pp 411--414

\bibitem[\protect\citeauthoryear{{Lodders}, {Palme}  \& {Gail}}{{Lodders}
  et~al.}{2009}]{2009LanB...4B...44L}
{Lodders} K.,  {Palme} H.,   {Gail} H.-P.,  2009, \mn@doi [Landolt
  B{\"o}rnstein] {10.1007/978-3-540-88055-4_34}, \href
  {http://adsabs.harvard.edu/abs/2009LanB...4B...44L} {p.~44}

\bibitem[\protect\citeauthoryear{{Mao} \& {Kaastra}}{{Mao} \&
  {Kaastra}}{2016}]{2016A&A...587A..84M}
{Mao} J.,  {Kaastra} J.,  2016, \mn@doi [\aap] {10.1051/0004-6361/201527568},
  \href {http://adsabs.harvard.edu/abs/2016A%26A...587A..84M} {587, A84}

\bibitem[\protect\citeauthoryear{{Mernier}, {de Plaa}, {Lovisari}, {Pinto},
  {Zhang}, {Kaastra}, {Werner}  \& {Simionescu}}{{Mernier}
  et~al.}{2015}]{2015A&A...575A..37M}
{Mernier} F.,  {de Plaa} J.,  {Lovisari} L.,  {Pinto} C.,  {Zhang} Y.-Y.,
  {Kaastra} J.~S.,  {Werner} N.,   {Simionescu} A.,  2015, \mn@doi [\aap]
  {10.1051/0004-6361/201425282}, \href
  {http://adsabs.harvard.edu/abs/2015A%26A...575A..37M} {575, A37}

\bibitem[\protect\citeauthoryear{{Mernier}, {de Plaa}, {Pinto}, {Kaastra},
  {Kosec}, {Zhang}, {Mao}  \& {Werner}}{{Mernier}
  et~al.}{2016a}]{2016A&A...592A.157M}
{Mernier} F.,  {de Plaa} J.,  {Pinto} C.,  {Kaastra} J.~S.,  {Kosec} P.,
  {Zhang} Y.-Y.,  {Mao} J.,   {Werner} N.,  2016a, \mn@doi [\aap]
  {10.1051/0004-6361/201527824}, \href
  {http://adsabs.harvard.edu/abs/2016A%26A...592A.157M} {592, A157}

\bibitem[\protect\citeauthoryear{{Mernier} et~al.,}{{Mernier}
  et~al.}{2016b}]{2016A&A...595A.126M}
{Mernier} F.,  et~al., 2016b, \mn@doi [\aap] {10.1051/0004-6361/201628765},
  \href {http://adsabs.harvard.edu/abs/2016A%26A...595A.126M} {595, A126}

\bibitem[\protect\citeauthoryear{{Mernier} et~al.,}{{Mernier}
  et~al.}{2017}]{2017A&A...603A..80M}
{Mernier} F.,  et~al., 2017, \mn@doi [\aap] {10.1051/0004-6361/201630075},
  \href {http://adsabs.harvard.edu/abs/2017A%26A...603A..80M} {603, A80}

\bibitem[\protect\citeauthoryear{{Mernier} et~al.,}{{Mernier}
  et~al.}{2018}]{2018MNRAS.478L.116M}
{Mernier} F.,  et~al., 2018, \mn@doi [\mnras] {10.1093/mnrasl/sly080}, \href
  {http://adsabs.harvard.edu/abs/2018MNRAS.478L.116M} {478, L116}

\bibitem[\protect\citeauthoryear{{Mewe}}{{Mewe}}{1972}]{1972A&A....20..215M}
{Mewe} R.,  1972, \aap, \href
  {http://adsabs.harvard.edu/abs/1972A%26A....20..215M} {20, 215}

\bibitem[\protect\citeauthoryear{{Mewe}, {Gronenschild}  \& {van den
  Oord}}{{Mewe} et~al.}{1985}]{1985A&AS...62..197M}
{Mewe} R.,  {Gronenschild} E.~H.~B.~M.,   {van den Oord} G.~H.~J.,  1985,
  \aaps, \href {http://adsabs.harvard.edu/abs/1985A%26AS...62..197M} {62, 197}

\bibitem[\protect\citeauthoryear{{Mewe}, {Lemen}  \& {van den Oord}}{{Mewe}
  et~al.}{1986}]{1986A&AS...65..511M}
{Mewe} R.,  {Lemen} J.~R.,   {van den Oord} G.~H.~J.,  1986, \aaps, \href
  {http://adsabs.harvard.edu/abs/1986A%26AS...65..511M} {65, 511}

\bibitem[\protect\citeauthoryear{{Mulchaey}, {Kasliwal}  \&
  {Kollmeier}}{{Mulchaey} et~al.}{2014}]{2014ApJ...780L..34M}
{Mulchaey} J.~S.,  {Kasliwal} M.~M.,   {Kollmeier} J.~A.,  2014, \mn@doi
  [\apjl] {10.1088/2041-8205/780/2/L34}, \href
  {http://adsabs.harvard.edu/abs/2014ApJ...780L..34M} {780, L34}

\bibitem[\protect\citeauthoryear{{Mushotzky}, {Loewenstein}, {Arnaud},
  {Tamura}, {Fukazawa}, {Matsushita}, {Kikuchi}  \& {Hatsukade}}{{Mushotzky}
  et~al.}{1996}]{1996ApJ...466..686M}
{Mushotzky} R.,  {Loewenstein} M.,  {Arnaud} K.~A.,  {Tamura} T.,  {Fukazawa}
  Y.,  {Matsushita} K.,  {Kikuchi} K.,   {Hatsukade} I.,  1996, \mn@doi [\apj]
  {10.1086/177541}, \href {http://adsabs.harvard.edu/abs/1996ApJ...466..686M}
  {466, 686}

\bibitem[\protect\citeauthoryear{{Nomoto}, {Kobayashi}  \& {Tominaga}}{{Nomoto}
  et~al.}{2013}]{2013ARA&A..51..457N}
{Nomoto} K.,  {Kobayashi} C.,   {Tominaga} N.,  2013, \mn@doi [\araa]
  {10.1146/annurev-astro-082812-140956}, \href
  {http://adsabs.harvard.edu/abs/2013ARA%26A..51..457N} {51, 457}

\bibitem[\protect\citeauthoryear{{Pinto} et~al.,}{{Pinto}
  et~al.}{2015}]{2015A&A...575A..38P}
{Pinto} C.,  et~al., 2015, \mn@doi [\aap] {10.1051/0004-6361/201425278}, \href
  {http://adsabs.harvard.edu/abs/2015A%26A...575A..38P} {575, A38}

\bibitem[\protect\citeauthoryear{{Sato}, {Tokoi}, {Matsushita}, {Ishisaki},
  {Yamasaki}, {Ishida}  \& {Ohashi}}{{Sato} et~al.}{2007}]{2007ApJ...667L..41S}
{Sato} K.,  {Tokoi} K.,  {Matsushita} K.,  {Ishisaki} Y.,  {Yamasaki} N.~Y.,
  {Ishida} M.,   {Ohashi} T.,  2007, \mn@doi [\apjl] {10.1086/522031}, \href
  {http://adsabs.harvard.edu/abs/2007ApJ...667L..41S} {667, L41}

\bibitem[\protect\citeauthoryear{{Schellenberger}, {Reiprich}, {Lovisari},
  {Nevalainen}  \& {David}}{{Schellenberger}
  et~al.}{2015}]{2015A&A...575A..30S}
{Schellenberger} G.,  {Reiprich} T.~H.,  {Lovisari} L.,  {Nevalainen} J.,
  {David} L.,  2015, \mn@doi [\aap] {10.1051/0004-6361/201424085}, \href
  {http://adsabs.harvard.edu/abs/2015A%26A...575A..30S} {575, A30}

\bibitem[\protect\citeauthoryear{{Simionescu}, {Werner}, {B{\"o}hringer},
  {Kaastra}, {Finoguenov}, {Br{\"u}ggen}  \& {Nulsen}}{{Simionescu}
  et~al.}{2009}]{2009A&A...493..409S}
{Simionescu} A.,  {Werner} N.,  {B{\"o}hringer} H.,  {Kaastra} J.~S.,
  {Finoguenov} A.,  {Br{\"u}ggen} M.,   {Nulsen} P.~E.~J.,  2009, \mn@doi
  [\aap] {10.1051/0004-6361:200810225}, \href
  {http://adsabs.harvard.edu/abs/2009A%26A...493..409S} {493, 409}

\bibitem[\protect\citeauthoryear{{Simionescu}, {Werner}, {Urban}, {Allen},
  {Ichinohe}  \& {Zhuravleva}}{{Simionescu} et~al.}{2015}]{2015ApJ...811L..25S}
{Simionescu} A.,  {Werner} N.,  {Urban} O.,  {Allen} S.~W.,  {Ichinohe} Y.,
  {Zhuravleva} I.,  2015, \mn@doi [\apjl] {10.1088/2041-8205/811/2/L25}, \href
  {http://adsabs.harvard.edu/abs/2015ApJ...811L..25S} {811, L25}

\bibitem[\protect\citeauthoryear{{Simionescu} et~al.,}{{Simionescu}
  et~al.}{2018}]{2018arXiv180600932S}
{Simionescu} A.,  et~al., 2018, preprint, \href
  {http://adsabs.harvard.edu/abs/2018arXiv180600932S} {} (\mn@eprint {arXiv}
  {1806.00932})

\bibitem[\protect\citeauthoryear{{Urdampilleta}, {Kaastra}  \&
  {Mehdipour}}{{Urdampilleta} et~al.}{2017}]{2017A&A...601A..85U}
{Urdampilleta} I.,  {Kaastra} J.~S.,   {Mehdipour} M.,  2017, \mn@doi [\aap]
  {10.1051/0004-6361/201630170}, \href
  {http://adsabs.harvard.edu/abs/2017A%26A...601A..85U} {601, A85}

\bibitem[\protect\citeauthoryear{{Werner}, {B{\"o}hringer}, {Kaastra}, {de
  Plaa}, {Simionescu}  \& {Vink}}{{Werner} et~al.}{2006}]{2006A&A...459..353W}
{Werner} N.,  {B{\"o}hringer} H.,  {Kaastra} J.~S.,  {de Plaa} J.,
  {Simionescu} A.,   {Vink} J.,  2006, \mn@doi [\aap]
  {10.1051/0004-6361:20065678}, \href
  {http://adsabs.harvard.edu/abs/2006A%26A...459..353W} {459, 353}

\bibitem[\protect\citeauthoryear{{Werner}, {Durret}, {Ohashi}, {Schindler}  \&
  {Wiersma}}{{Werner} et~al.}{2008}]{2008SSRv..134..337W}
{Werner} N.,  {Durret} F.,  {Ohashi} T.,  {Schindler} S.,   {Wiersma} R.~P.~C.,
   2008, \mn@doi [\ssr] {10.1007/s11214-008-9320-9}, \href
  {http://adsabs.harvard.edu/abs/2008SSRv..134..337W} {134, 337}

\bibitem[\protect\citeauthoryear{{de Plaa}, {Werner}, {Bleeker}, {Vink},
  {Kaastra}  \& {M{\'e}ndez}}{{de Plaa} et~al.}{2007}]{2007A&A...465..345D}
{de Plaa} J.,  {Werner} N.,  {Bleeker} J.~A.~M.,  {Vink} J.,  {Kaastra} J.~S.,
   {M{\'e}ndez} M.,  2007, \mn@doi [\aap] {10.1051/0004-6361:20066382}, \href
  {http://adsabs.harvard.edu/abs/2007A%26A...465..345D} {465, 345}

\bibitem[\protect\citeauthoryear{{de Plaa} et~al.,}{{de Plaa}
  et~al.}{2017}]{2017A&A...607A..98D}
{de Plaa} J.,  et~al., 2017, \mn@doi [\aap] {10.1051/0004-6361/201629926},
  \href {http://adsabs.harvard.edu/abs/2017A%26A...607A..98D} {607, A98}

\makeatother
\end{thebibliography}




%
%


\bsp	
\label{lastpage}
\end{document}